\documentclass[11pt]{article}

\usepackage[cmtip,arrow]{xy}
\usepackage{pb-diagram,pb-xy}
\usepackage{amssymb,amsopn}

\newlength{\vshift}
\newlength{\hshift}
\def\la{\lambda}

\def\a{\alpha}

\def\ds{\stackrel{\star}{,}}

\def\p{\partial}

\def\KK{K^{m}_{\ \ l}K^{*nl}}
\def\KKt{K^{ab}K_{ab}}
\def\KKtc{K^{*cd}K_{cd}^*}

\def\tr{{\rm Tr}}

\def\d{\delta}
\def\G{\Gamma}
\def\S{\Sigma}

\begin{document}

\begin{titlepage}

$\,$

\vspace{1.5cm}
\begin{center}

{\LARGE{\bf 
Twisted SUSY: twisted symmetry versus renormalizability
}}

\vspace*{1.3cm}

{{\bf Marija Dimitrijevi\' c, Biljana Nikoli\' c and\\
Voja Radovanovi\' c }}

\vspace*{1cm}

University of Belgrade, Faculty of Physics\\
Studentski trg 12, 11000 Beograd, Serbia \\[1em]

\end{center}

\vspace*{2cm}

\begin{abstract}

We discuss a deformation of superspace based on a hermitian
twist. The twist implies a $\star$-product that is
noncommutative, hermitian and finite when expanded in power series
of the deformation parameter. The Leibniz rule for the twisted
SUSY transformations is deformed. A minimal deformation
of the Wess-Zumino action is proposed and its renormalizability 
properties are discussed. There is no tadpole contribution, but 
the two-point function diverges. We speculate that the deformed Leibniz 
rule, or more generally the twisted 
symmetry, interferes with renormalizability properties of the model. We discuss
different possibilities to render a renormalizable model.

\end{abstract}
\vspace*{1cm}



\vspace*{1cm}
\quad\scriptsize{eMail:
dmarija,biljana,rvoja@ipb.ac.rs}
\vfill

\end{titlepage}\vskip.2cm

\newpage
\setcounter{page}{1}
\newcommand{\Section}[1]{\setcounter{equation}{0}\section{#1}}
\renewcommand{\theequation}{\arabic{section}.\arabic{equation}}

\section{Introduction}

It is well known that Quantum Field Theory encounters
problems at high energies and short distances. This
suggests that the structure of space-time has to be modified
at these scales. One possibility to modify the structure of space-time is to
deform the usual commutation relations between coordinates; this gives 
a noncommutative (NC) space \cite{noncommspace}. Different models of 
noncommutativity were discussed in the literature, see 
\cite{NCbooks}, \cite{NCbookMi} and \cite{NCreview} for references. A version
of Standard Model on the canonically deformed space-time was 
constructed in \cite{NCSM} and its renormalizability properties were
discussed in \cite{NCSMRenorm}. Renormalizability of different
noncommutative field theory models was discussed in \cite{NCRenorm}.

A natural further step is modification of the superspace
and introduction of non\-(anti)\-commutativity. A strong motivation
for this comes from string theory. Namely, it was discovered that
a noncommutative superspace can arise when a 
superstring moves in a constant gravitino or graviphoton background 
\cite{Seiberg}, \cite{NACStrings}. Since that discovery there has been a lot
of work on this subject and different ways of deforming superspace 
have been discussed. Here we mention some of them.

The authors of \cite{luksusy}
combine SUSY with the $\kappa$-deformation
of space-time, while in \cite{MWsusy} SUSY is combined with the
canonical deformation of space-time. In \cite{Seiberg}
a version of non(anti)commutative superspace is defined and analyzed.  
The anticommutation relations between fermionic coordinates are
modified in the following way
\begin{equation}
\{ \theta^\alpha \ds \theta^\beta\} =C^{\alpha\beta} , \quad \{
\bar{\theta}_{\dot\alpha} \ds \bar{\theta}_{\dot\beta}\} = \{
\theta^\alpha \ds \bar{\theta}_{\dot\alpha}\} = 0\ ,
\label{eucliddef}
\end{equation}
where $C^{\alpha\beta} = C^{\beta\alpha}$ is a complex, constant
symmetric matrix. This deformation is well defined only when undotted 
and dotted spinors are not related by the
usual complex conjugation. The notion of chirality is preserved in this model, i.e.
the deformed product of two chiral superfields is again a chiral
superfield. On the other hand, one half of ${\cal N}=1$ supersymmetry is broken and this is
the so-called ${\cal N}=1/2$ supersymmetry. Another type of deformation is
introduced in \cite{Ferrara} and \cite{D-def-us}. 
There the product of two chiral
superfields is not a chiral superfield but the model is invariant
under the full supersymmetry. Renormalizability of different models
(both scalar and gauge theories) has been discussed in \cite{RenWZ}, 
\cite{Penati}, \cite{RenYM} and \cite{D-def-us}. The twist approach to
nonanticommutativity was discussed in \cite{SUSYtwist}.

In our previous paper \cite{miSUSY} we introduced a hermitian deformation
of the usual superspace. The non(anti)commutative 
deformation  was introduced via the twist
\begin{equation}
{\cal F} = e^{\frac{1}{2}C^{\alpha\beta}\p_\alpha \otimes\p_\beta
+ \frac{1}{2}\bar{C}_{\dot{\alpha}\dot{\beta}}
\bar{\p}^{\dot{\alpha}}\otimes\bar{\p}^{\dot{\beta}} } .\label{intro-twist}
\end{equation}
Here $C^{\alpha\beta} = C^{\beta\alpha}$ is a
complex constant matrix, $\bar{C}^{\dot{\alpha}\dot{\beta}}$ its complex
conjugate and $\p_\alpha = \frac{\p}{\p \theta^\alpha}$ are
fermionic partial derivatives. The twist (\ref{intro-twist}) is
hermitian under the usual complex conjugation. Due to this choice
of the twist, the coproduct of the SUSY transformations becomes
deformed, leading to the deformed Leibniz rule. The inverse of
(\ref{intro-twist}) defines the $\star$-product. It is obvious
that the $\star$-product of two chiral fields will not be a chiral
field. Therefore we have to use the method of projectors to 
decompose the $\star$-products of fields into their irreducible components.
Collecting the terms invariant under the twisted SUSY transformations we construct
the deformed Wess-Zumino action.

Being interested in implications of the twisted symmetry on
renormalizability properties, in this paper we calculate
the divergent part of the one-loop effective action. More precisely, we calculate divergent parts of the one-point and the two-point functions. The plan of the 
paper is as follows: In the next section we
summarize the most important properties of our model, more details
of the construction are given in \cite{miSUSY}. In Section 3 we describe 
the method we use to calculate divergent
parts of the $n$-point Green functions: the background field method and 
the supergraph technique. In Sections 4 the tadpole diagram and the divergent part of the two-point
function are calculated. Finally, we discuss renormalizability of the 
model. We give some comments and compare our results with the results already
present in the literature. Some details of our calculations are
presented in the Appendix.

\section{Construction of the model}

There are different ways to realize a noncommutative and/or a nonanticommutative
space and to formulate
a physical model on it, see \cite{NCbooks} and \cite{NCreview}. We shall 
follow the approach of \cite{NCbookMi} and \cite{miSUSY}.

Let us first fix the notation and the conventions which we use.
The superspace is generated by supercoordinates $x^{m}$, $\theta^{\alpha}$
and $\bar{\theta}_{\dot{\alpha}}$ which fulfill
\begin{eqnarray}
\lbrack x^m, x^n \rbrack = \lbrack x^m,
\theta^\alpha \rbrack = \lbrack x^m, \bar{\theta}_{\dot\alpha} \rbrack = 0 ,\quad
\{ \theta^\alpha , \theta^\beta\} = \{ \bar{\theta}_{\dot\alpha}
, \bar{\theta}_{\dot\beta}\} = \{ \theta^\alpha ,
\bar{\theta}_{\dot\alpha}\} = 0 , \label{undefsupsp}
\end{eqnarray}
with $m=0,\dots ,3$ and $\alpha, \beta =1,2$. To $x^m$ we refer as bosonic and to
$\theta^\alpha$ and $\bar{\theta}_{\dot\alpha}$ we refer as
fermionic coordinates. We work in Minkowski space-time with the
metric $(-,+,+,+)$ and $x^m x_m = -(x^0)^2 + (x^1)^2 + (x^2)^2 +
(x^3)^2$.

A general superfield ${\rm F}(x,\theta, \bar{\theta})$ can be expanded
in powers of $\theta$ and $\bar{\theta}$,
\begin{eqnarray}
{\rm F}(x, \theta, \bar{\theta}) &=&\hspace*{-2mm} f(x) + \theta\phi(x)
+ \bar{\theta}\bar{\chi}(x)
+ \theta\theta m(x) + \bar{\theta}\bar{\theta} n(x) + \theta\sigma^m\bar{\theta}v_m(x)\nonumber\\
&& + \theta\theta\bar{\theta}\bar{\lambda}(x) + \bar{\theta}\bar{\theta}\theta\varphi(x)
+ \theta\theta\bar{\theta}\bar{\theta} d(x) .\label{F}
\end{eqnarray}
Under the infinitesimal ${\cal N} = 1$ SUSY transformations it transforms as
\begin{equation}
\delta_\xi {\rm F} = \big(\xi Q + \bar{\xi}\bar{Q} \big) {\rm F}, \label{susytr}
\end{equation}
where $\xi^{\alpha}$ and $\bar{\xi}_{\dot{\alpha}}$ are constant
anticommuting parameters and the SUSY generators $Q^{\alpha}$ and
$\bar{Q}_{\dot\alpha}$ are given by,
\begin{eqnarray}
Q_\alpha = \p_\alpha
- i\sigma^m_{\ \alpha\dot{\alpha}}\bar{\theta}^{\dot{\alpha}}\p_m, \quad \bar{Q}_{\dot{\alpha}} = -\bar{\p}_{\dot{\alpha}} + i\theta^\alpha \sigma^m_{\
\alpha\dot{\alpha}}\p_m
.\label{q,barq}
\end{eqnarray}
Transformations (\ref{susytr}) close in the algebra
\begin{equation}
[\delta_\xi, \delta_\eta] = -2i(\eta\sigma^m \bar{\xi} 
- \xi\sigma^m \bar{\eta})\p_m. \label{xietaalg}
\end{equation}
The product of two superfields is a superfield again; its transformation
law is given by
\begin{eqnarray}
\delta_\xi ({\rm F}\cdot {\rm G}) &=& \big(\xi Q + \bar{\xi}\bar{Q} \big) 
({\rm F}\cdot {\rm G}), \nonumber\\
&=& (\delta_\xi {\rm F})\cdot {\rm G} + {\rm F}\cdot(\delta_\xi {\rm G}).
\label{undefLrule}
\end{eqnarray}
The last line is the undeformed Leibniz rule for the infinitesimal
SUSY transformation $\delta_\xi$.

Nonanticommutativity is introduced following the 
twist approach \cite{NCbookMi}. For the twist $\cal{F}$ we choose
\begin{equation}
{\cal F} = e^{\frac{1}{2}C^{\alpha\beta}\p_\alpha \otimes\p_\beta
+ \frac{1}{2}\bar{C}_{\dot{\alpha}\dot{\beta}}\bar{\p}^{\dot{\alpha}}\otimes\bar{\p}^{\dot{\beta}} } ,\label{twist}
\end{equation}
with the complex constant matrix $C^{\alpha\beta} = C^{\beta\alpha}$. Note that
$C^{\alpha\beta}$ and $\bar{C}^{\dot{\alpha}\dot{\beta}}$ are
related by the usual complex conjugation. It can be shown that
the twist (\ref{twist}) satisfies all necessary requirements \cite{chpr}.

The inverse of the twist (\ref{twist})
\begin{equation}
{\cal F}^{-1} = e^{-\frac{1}{2}C^{\alpha\beta}\p_\alpha
\otimes\p_\beta -
\frac{1}{2}\bar{C}_{\dot{\alpha}\dot{\beta}}\bar{\p}^{\dot{\alpha}}\otimes\bar{\p}^{\dot{\beta}}
} ,\label{invtwist}
\end{equation}
defines a new product in the algebra of superfields called the 
$\star$-product. For two arbitrary
superfields ${\rm F}$ and ${\rm G}$ the $\star$-product is defined as follows
\begin{eqnarray}
{\rm F}\star {\rm G} &=& \mu_\star \{ {\rm F}\otimes {\rm G} \} \nonumber\\
&=& \mu \{ {\cal F}^{-1}\, {\rm F}\otimes {\rm G}\} \nonumber\\
&=& \mu \{ e^{-\frac{1}{2}C^{\alpha\beta}\p_\alpha \otimes\p_\beta
-
\frac{1}{2}\bar{C}_{\dot{\alpha}\dot{\beta}}\bar{\p}^{\dot{\alpha}}
\otimes\bar{\p}^{\dot{\beta}}} {\rm F}\otimes {\rm G} \} \nonumber\\
&=& {\rm F}\cdot {\rm G} - \frac{1}{2}(-1)^{|{\rm F}|}C^{\alpha\beta}
(\p_\alpha {\rm F})\cdot(\p_\beta {\rm G}) -
\frac{1}{2}(-1)^{|{\rm F}|}\bar{C}_{\dot{\alpha}\dot{\beta}}
(\bar{\p}^{\dot{\alpha}} {\rm F})
(\bar{\p}^{\dot{\beta}} {\rm G})\nonumber\\
&&- \frac{1}{8}C^{\alpha\beta}C^{\gamma\delta}(\p_\alpha\p_\gamma {\rm F})
\cdot(\p_\beta\p_\delta {\rm G})
- \frac{1}{8}\bar{C}_{\dot{\alpha}\dot{\beta}}\bar{C}_{\dot{\gamma}\dot{\delta}}
(\bar{\p}^{\dot{\alpha}}\bar{\p}^{\dot{\gamma}} {\rm F})
(\bar{\p}^{\dot{\beta}}\bar{\p}^{\dot{\delta}} {\rm G}) \nonumber\\
&&- \frac{1}{4}C^{\alpha\beta}\bar{C}_{\dot{\alpha}\dot{\beta}}
(\p_\alpha\bar{\p}^{\dot{\alpha}}{\rm F})
(\p_\beta\bar{\p}^{\dot{\beta}} {\rm G}) \nonumber\\
&& + \frac{1}{16}(-1)^{|{\rm F}|}
C^{\alpha\beta}C^{\gamma\delta}\bar{C}_{\dot{\alpha}\dot{\beta}}
(\p_\alpha\p_\gamma\bar{\p}^{\dot{\alpha}} {\rm F})
(\p_\beta\p_\delta\bar{\p}^{\dot{\beta}} {\rm G}) \nonumber\\
&& + \frac{1}{16}(-1)^{|{\rm F}|}C^{\alpha\beta}\bar{C}_{\dot{\alpha}
\dot{\beta}} \bar{C}_{\dot{\gamma}\dot{\delta}}
(\p_\alpha\bar{\p}^{\dot{\alpha}}\bar{\p}^{\dot{\gamma}} {\rm F})
(\p_\beta\bar{\p}^{\dot{\beta}}\bar{\p}^{\dot{\delta}} {\rm G})
\nonumber\\
&& + \frac{1}{64}C^{\alpha\beta}C^{\gamma\delta}
\bar{C}_{\dot{\alpha}\dot{\beta}}\bar{C}_{\dot{\gamma}\dot{\delta}}
(\p_\alpha\p_\gamma\bar{\p}^{\dot{\alpha}}\bar{\p}^{\dot{\gamma}}
{\rm F})(\p_\beta\p_\delta \bar{\p}^{\dot{\beta}}\bar{\p}^{\dot{\delta}}
{\rm G}), \label{star}
\end{eqnarray}
where $|{\rm F}| = 1$ if ${\rm F}$ is odd (fermionic) and $|{\rm F}|=0$ 
if ${\rm F}$ is even 
(bosonic) and the pointwise multiplication $\mu$ is the bilinear map from the tensor product to the space of
superfields (functions). The definition of the multiplication $\mu_\star$
 is given in the second line.
No higher powers of $C^{\alpha\beta}$ and
$\bar{C}_{\dot{\alpha}\dot{\beta}}$ appear since the derivatives
$\p_\alpha$ and $\bar{\p}^{\dot{\alpha}}$ are Grassmanian.
Expansion of the $\star$-product (\ref{star}) ends after the fourth 
order in the deformation parameter. This $\star$-product is different from the Moyal-Weyl 
$\star$-product \cite{mw} where the expansion in powers of
the deformation parameter leads to an infinite power series. One should
also note that the $\star$-product (\ref{star}) is hermitian,
\begin{equation}
({\rm F}\star {\rm G})^* = {\rm G}^* \star {\rm F}^* , \label{complconj}
\end{equation}
where $*$ denotes the usual complex conjugation.

The $\star$-product (\ref{star}) implies
\begin{eqnarray}
\{ \theta^\alpha \ds \theta^\beta \} &=& C^{\alpha\beta}, \quad \{
\bar{\theta}_{\dot\alpha}\ds \bar{\theta}_{\dot\beta}\} =
\bar{C}_{\dot{\alpha}\dot{\beta}}, \quad \{ \theta^\alpha \ds
\bar{\theta}_{\dot\alpha} \} = 0 ,\nonumber \\
\lbrack x^m \ds x^n \rbrack &=& 0 , \quad [x^m \ds \theta^\alpha ]
= 0, \quad [x^m \ds \bar{\theta}_{\dot\alpha} ] = 0 .
\label{thetastar}
\end{eqnarray}
Relations (\ref{thetastar}) enable us to define the deformed
superspace or "non\-anti\-commutative
superspace". It is generated by the usual bosonic and fermionic
coordinates (\ref{undefsupsp}) while the deformation is contained
in the new product (\ref{star}).

The next step is to apply the twist (\ref{twist}) to the Hopf algebra of
SUSY transformations. We will not give details here, they can be
found in \cite{miSUSY}. We just state the most important results.

The deformed infinitesimal SUSY transformation is defined in the following way
\begin{equation}
\delta^\star_\xi {\rm F} = \big(\xi Q + \bar{\xi}\bar{Q} \big) {\rm F}.
\label{defsusytr}
\end{equation}
The twist (\ref{twist}) leads to a deformed Leibniz rule for the 
deformed SUSY transformations (\ref{defsusytr}). This ensures that the
$\star$-product of two superfields is again a superfield. Its
transformation law is given by
\begin{eqnarray}
\delta^\star_\xi ({\rm F}\star {\rm G}) &=& \big(\xi Q 
+ \bar{\xi}\bar{Q} \big) ({\rm F}\star {\rm G}), \label{deftrlaw}\\
&=&
(\delta^\star_\xi {\rm F}) \star {\rm G} 
+ {\rm F} \star (\delta^\star_\xi {\rm G}) \nonumber\\
&& + \frac{i}{2}C^{\alpha\beta}\Big( \bar{\xi}^{\dot{\gamma}}
\sigma^m_{\ \alpha\dot{\gamma}}(\p_m {\rm F})\star (\p_\beta {\rm G}) +
(\p_\alpha {\rm F})\star \bar{\xi}^{\dot{\gamma}}
\sigma^m_{\ \beta\dot{\gamma}}(\p_m {\rm G}) \Big) \label{defLpravilo}\\
&& -\frac{i}{2}\bar{C}_{\dot{\alpha}\dot{\beta}} \Big(
\xi^\alpha\sigma^m_{\
\alpha\dot{\gamma}}\varepsilon^{\dot{\gamma}\dot{\alpha}}(\p_m {\rm F})
\star (\bar{\p}^{\dot{\beta}} {\rm G}) + (\bar{\p}^{\dot{\alpha}} {\rm F})
\star \xi^\alpha\sigma^m_{\ \alpha\dot{\gamma}}
\varepsilon^{\dot{\gamma}\dot{\beta}}(\p_m {\rm G}) \Big) .\nonumber
\end{eqnarray}
Note that we have to enlarge the algebra (\ref{xietaalg}) by introducing the 
fermionic derivatives $\p_\alpha$ and $\bar{\p}_{\dot{\alpha}}$. Since 
these derivatives commute with the generators of Poincar\' e algebra 
$\p_m$ and $M_{mn}$, the super Poincar\' e algebra does not change. 
Especially, the Leibniz rule for $\p_m$ and $M_{mn}$ does not change.

Being interested in a deformation of the Wess-Zumino model, we
need to analyze properties of the $\star$-products of chiral fields.
A chiral field $\Phi$ fulfills $\bar{D}_{\dot{\alpha}}\Phi =0$,
with the supercovariant derivative $\bar{D}_{\dot{\alpha}} = -\bar{\p}_{\dot{\alpha}} -
i\theta^\alpha \sigma^m_{\ \alpha\dot{\alpha}}\p_m$. In terms of component fields the chiral
superfield $\Phi$ is given by
\begin{eqnarray}
\Phi(x, \theta, \bar{\theta}) &=& A(x) 
+ \sqrt{2}\theta^\alpha\psi_\alpha(x)
+ \theta\theta F(x) + i\theta\sigma^l\bar{\theta}\p_l A(x) \nonumber\\
&& -\frac{i}{\sqrt{2}}\theta\theta\p_m\psi^\alpha(x)
\sigma^m_{\ \alpha\dot{\alpha}}\bar{\theta}^{\dot{\alpha}}
+ \frac{1}{4}\theta\theta\bar{\theta}\bar{\theta}\Box A(x).\label{chiral}
\end{eqnarray}
It is easy to calculate the $\star$-product of two chiral fields from
(\ref{star}). It is given by
\begin{eqnarray}
\Phi\star\Phi &=& A^2 - \frac{C^2}{2}F^2 +
\frac{1}{4}C^{\alpha\beta}\bar{C}^{\dot{\alpha}\dot{\beta}}
\sigma^m_{\ \alpha\dot{\alpha}}\sigma^l_{\ \beta\dot{\beta}}(\p_m
A)(\p_l A)
+\frac{1}{64}C^2\bar{C}^2 (\Box A)^2\nonumber\\
&& + \theta^\alpha\Big( 2\sqrt{2}\psi_\alpha A 
-\frac{1}{\sqrt{2}}C^{\gamma\beta}\bar{C}^{\dot{\alpha}\dot{\beta}}
\varepsilon_{\gamma\alpha}
(\p_m\psi^\rho)\sigma^m_{\ \rho\dot{\beta}}\sigma^l_{\ \beta\dot{\alpha}}
(\p_l A) \Big)
\nonumber\\
&& -\frac{i}{\sqrt{2}}C^2 \bar{\theta}_{\dot{\alpha}}
\bar{\sigma}^{m\dot{\alpha}\alpha}(\p_m\psi_\alpha)F
+ \theta\theta \Big( 2AF - \psi\psi\Big) \nonumber\\
&& + \bar{\theta}\bar{\theta}\Big( -\frac{C^2}{4}\big( F\Box A
- \frac{1}{2}(\p_m\psi)\sigma^m\bar{\sigma}^l(\p_l\psi) \big) \Big) 
\nonumber\\
&& + i\theta\sigma^m\bar{\theta}\Big( (\p_m A^2) 
+ \frac{1}{4}C^{\alpha\beta}\bar{C}^{\dot{\alpha}\dot{\beta}}
\sigma_{m\alpha\dot{\alpha}}\sigma^l_{\ \beta\dot{\beta}}(\Box A)
(\p_l A)\Big) \nonumber\\
&& +  i\sqrt{2}\theta\theta\bar{\theta}_{\dot{\alpha}}
\bar{\sigma}^{m\dot{\alpha}\alpha}
\big( \p_m(\psi_\alpha A)\big) + \frac{1}{4}\theta\theta\bar{\theta}
\bar{\theta} (\Box A^2) , \label{phistarphi}
\end{eqnarray}
where $C^2 =
C^{\alpha\beta}C^{\gamma\delta}\varepsilon_{\alpha\gamma}\varepsilon_{\beta\delta}$
and $\bar{C}^2 =
\bar{C}_{\dot{\alpha}\dot{\beta}}\bar{C}_{\dot{\gamma}\dot{\delta}}
\varepsilon^{\dot{\alpha}\dot{\gamma}}\varepsilon^{\dot{\beta}\dot{\delta}}$.
One sees that due to the $\bar{\theta}$-term and the 
$\bar{\theta}\bar{\theta}$-term (\ref{phistarphi}) is not a chiral 
field. But, in order to write an action invariant under the deformed
SUSY transformations (\ref{defsusytr}) we need to preserve the
notion of chirality. This can be done in different ways. One possibility 
is to use a different $\star$-product, the one which preserves
chirality \cite{Seiberg}. However, chirality-preserving
$\star$-product implies working in a space where
$\bar{\theta} \neq (\theta)^*$. Since we want to work in Minkowski
space-time and keep the usual complex conjugation, we use the
$\star$-product (\ref{star}) and decompose the $\star$-products of superfields
into their irreducible components using the projectors defined in
\cite{wessbook}. In that way (\ref{phistarphi}) becomes
\begin{equation}
\Phi\star \Phi = P_1(\Phi\star \Phi) + P_2 (\Phi\star \Phi) 
+ P_T(\Phi\star \Phi), 
\end{equation}
with antichiral, chiral and transversal projectors given by
\begin{eqnarray}
P_1 &=& \frac{1}{16} \frac{D^2 \bar{D}^2}{\Box}, \label{P1}\\
P_2 &=& \frac{1}{16} \frac{\bar{D}^2 D^2 }{\Box}, \label{P2}\\
P_T &=& -\frac{1}{8} \frac{D \bar{D}^2 D}{\Box}. \label{PT}
\end{eqnarray}

Finally, the deformed Wess-Zumino action is constructed requiring 
that the action is invariant under the
deformed SUSY transformations (\ref{defsusytr}) and that in the
commutative limit it reduces to the undeformed Wess-Zumino action. In 
addition, we require that deformation is minimal: We deform only those 
terms that are present in the commutative Wess-Zumino 
model. We do not, for the time being, add the terms whose commutative 
limit is zero.

Taking these requirements into account we propose the following action
\begin{eqnarray}
S &=& \int {\mbox{d}}^4 x\hspace{1mm} \Big{\{} \Phi^+\star\Phi\Big|_{\theta\theta\bar\theta\bar\theta}
+ \Big[ \frac{m}{2}P_2\big( \Phi\star\Phi\big)
\Big|_{\theta\theta} \nonumber\\
&& + \frac{\lambda}{3}P_2\Big( \Phi\star
P_2\big( \Phi\star\Phi\big)\Big) \Big |_{\theta\theta} + {\mbox{
c.c }} \Big] \Big{\}}, \label{S}
\end{eqnarray}
where $m$ and $\lambda$ are real constants. To rewrite (\ref{S}) 
in terms of component fields and as compact as possible,
we introduce the following notation
\begin{eqnarray}
C_{\alpha\beta}&=& K_{ab}(\sigma^{ab}\varepsilon)_{\alpha\beta},\label{n1}\\
\bar{C}_{\dot{\alpha}\dot{\beta}}&=&
K^*_{ab}(\varepsilon\bar{\sigma}^{ab})_{\dot{\alpha}\dot{\beta}},\label{n2}
\end{eqnarray}
where $K_{ab}=-K_{ba}$ is an antisymmetric self-dual complex
constant matrix. Then we have
\begin{eqnarray}
C^2 = 2K_{ab}K^{ab}, && \bar{C}^2 = 2K^*_{ab}K^{*ab}, \quad
K^{ab}K^*_{ab}= 0.\label{n4} \\
K^*_{cd}K_{ab}\big( \sigma^n\bar{\sigma}^{cd}\bar{\sigma}^m
\sigma^{ab}\big)_\alpha^{\ \beta} &=&
-4\delta^\beta_\alpha K^{ma}K^{*n}_{\ \ a} + 8 K^{ma}K^{*nb}(\sigma_{ba})_\alpha^{\ \beta} ,\label{n5}\\
C^{\alpha\beta}\bar{C}^{\dot{\alpha}\dot{\beta}} \sigma^m_{\
\alpha\dot{\alpha}}\sigma^l_{\ \beta\dot{\beta}} &=& 8K^{am}K^{*\
l}_a. \label{n6}
\end{eqnarray}
Using these formulas and expanding (\ref{S}) in component fields we obtain
\begin{eqnarray}
S &=& \int {\mbox{d}}^4 x\hspace{1mm} \Big{\{}
\Phi^+\star\Phi\Big|_{\theta\theta\bar\theta\bar\theta}\nonumber\\
&& + \Big[ \frac{m}{2}P_2\big( \Phi\star\Phi\big)
\Big|_{\theta\theta} + \frac{\lambda}{3}P_2\Big( \Phi\star P_2\big(
\Phi\star\Phi\big)\Big) \Big |_{\theta\theta} + {\mbox{ c.c }} \Big] 
\Big{\}}\nonumber\\
&=&
\int {\mbox{d}}^4 x\hspace{1mm} \Big{\{}
A^*\Box A + i(\p_m\bar\psi)\bar{\sigma}^m\psi + F^*F\nonumber\\
&& + \Big[ \frac{m}{2}\big( 2AF - \psi\psi \big) 
+\lambda\big( FA^2 - A\psi\psi  \big)\nonumber\\
&& -\frac{\lambda}{3}\Big(K^{m}_{\ a}K^{*na}\psi(\p_n\psi) - 2K^m_{\
a}K^{*n}_{\ \ b}(\p_n\psi)\sigma^{ba}\psi\Big) (\p_m A)\nonumber\\
&& -\frac{\lambda}{12}K^{mn}K_{mn}F^3
+\frac{\la}{6}\KK F(\p_m A)(\p_n A) \label{Sincompfields}\\
&& +\frac{\la}{3}\KK F\frac{1}{\Box}\p_m\Big( (\p_n A)\Box
A\Big) +\frac{\la}{192}\KKt\KKtc  F(\Box A)^2  + c.c \Big]
\Big{\}}. \nonumber
\end{eqnarray}
Partial integration was used to rewrite some of the terms in
(\ref{Sincompfields}) in a more compact way. Note that this it the 
complete action; there are no higher order terms in the deformation parameter
$K^{ab}$. However, for simplicity in the following sections we shall keep only 
terms up to second order in the deformation parameter.

\section{One-loop effective action}

In this section we look at the quantum properties of our model. We 
calculate the one-loop divergent part of the
one-point and the two-point functions up to second order in the 
deformation parameter.
We use the background field method, dimensional regularization and
the supergraph technique. The supergraph
technique significantly simplifies calculations. However,
we cannot directly apply this technique since our action 
(\ref{Sincompfields}) is not written as an integral over the whole 
superspace and in terms of the chiral field $\Phi$ and its derivatives. 
This is a consequence of the particular deformation
(\ref{F}) and differs from \cite{D-def-us}.

In order to be able to use the supergraph technique we notice 
the following: From
(\ref{chiral}), see also \cite{wessbook}, it follows that the fields
$A$, $\psi$ and $F$ can be written as
\begin{equation}
A=\Phi|_{\theta, \bar{\theta} =0}, \quad
\psi_{\alpha}=\frac{1}{\sqrt{2}}D_{\alpha}\Phi|_{\theta, \bar{\theta} 
=0}, \quad
F=-\frac{1}{4}D^{2}\Phi|_{\theta, \bar{\theta} =0}. \label{cfieldsPhi}
\end{equation}
Inserting this in (\ref{Sincompfields}) we obtain
\begin{eqnarray}
S &=& \int {\mbox{d}}^8 z\hspace{1mm} \Big{\{} \Phi^+\Phi +
\Big[ -\frac{m}{8}\Phi\frac{D^2}{\Box}\Phi -\lambda\Phi^2\frac{D^2}{12\Box}
\Phi\nonumber\\
&& +\lambda\theta\theta\bar{\theta}\bar{\theta} \Big( 
\frac{1}{768} K^{mn}K_{mn}(D^{2}\Phi)^{3} \nonumber \\
& &-\frac{1}{6}\left(K^{m}_{\ a}K^{*na}
(D^{\alpha}\Phi)(\partial_{n}D_{\alpha}\Phi)- 2K^{m}_{\ a}
 K^{*n}_{\ b}
(\partial_{n}D^{\alpha}\Phi)(\sigma^{ba})_{\alpha}^{\ \beta}D_{\beta}\Phi\right)(\partial_{m}\Phi)\nonumber\\
& & -
\frac{1}{24} K^{m}_{\ a}K^{*na}(D^{2}\Phi)(\partial_{m}\Phi)
 (\partial_{n}\Phi) \nonumber \\ 
& & -\frac{1}{12}K^{m}_{\ a} K^{*na}(D^{2}\Phi)\frac{1}{\Box}
 \partial_{m}\big( (\partial_{n}\Phi)(\square\Phi) \big) \Big)
 +c.c. \Big] \Big{\}}, \label{SGaction}
\end{eqnarray}
with $f(x)\frac{1}{\Box}g(x) = 
f(x)\int {\mbox{d}}^4 y\hspace{1mm} G(x-y)g(y)$.
Notice that two spurion fields 
\begin{equation}
U_{(1)\ \ ab}^{mn} = K^{m}_{\ a} K^{*n}_{\ b}\theta\theta\bar{\theta}\bar{\theta}, \quad
U_{(2)} = K^{mn}K_{mn}\theta\theta\bar{\theta}\bar{\theta}
\end{equation}
appear in (\ref{SGaction}). This is a consequence of rewriting the action (\ref{Sincompfields}) as an integral over the whole superspace.

Now we can start the machinery of the background field method. First we 
split the chiral and antichiral superfields into their classical and 
quantum parts
\begin{eqnarray}
\Phi\to \Phi+ \Phi_q ,\quad \Phi^+ \to \Phi^{+}+\Phi^{+}_q
\end{eqnarray}
and integrate over the quantum superfields in the path integral. Since 
$\Phi_q$ and $\Phi_q^{+}$ are chiral and antichiral fields, they are 
constrained by
$$\bar D_{\dot\a}\Phi_q= D_{\a}\Phi_q^+=0 .$$
To simplify the supergraph technique we introduce the
unconstrained superfields $\S$ and $\S^+$,
\begin{equation}
\Phi_{q} = -\frac{1}{4}\bar{D}^2\Sigma ,\quad
\Phi_{q}^{+} = -\frac{1}{4}D^2\Sigma^{+}\ .\label{abel-gaug-tr}
\end{equation}
Note that we do not express the background superfields $\Phi$ and
$\Phi^+$ in terms of $\S$ and $\S^+$, only the quantum parts
$\Phi_q$ and $\Phi_q^+$. After the integration of quantum
superfields, the result is expressed in terms of the
(anti)chiral superfields. This is a big advantage of the
background field method and of the supergraph technique. The unconstrained 
superfields are determined up to a gauge transformation
\begin{equation}
\S \to \S+\bar
D_{\dot\a}\bar\Lambda^{\dot\a}, \quad 
\S^+ \to \S^++D^{\a}\Lambda_{\a} , \label{gf1}
\end{equation}
with the gauge parameter $\Lambda$. This additional symmetry has to be fixed,
so we add a gauge-fixing term to the action. For the
gauge functions we choose
\begin{equation}
\chi_\a = D_\a\S ,\quad \bar \chi_{\dot\a} = \bar
D_{\dot\a}\S^+ \ . \label{gf2}
\end{equation}
The product $\d(\chi)\d(\bar\chi)$ in the path integral is
averaged by the weight $e^{-i\xi \int d^8z\bar{f}Mf}$:
\begin{equation}
\int {\mbox{d}}f {\mbox{d}}\bar f \hspace{1mm}
\d(\chi_\a-f_\a)\d(\bar\chi^{\dot\a}-\bar f^{\dot\a})e^{-i\xi\int
d^8 z\bar f^{\dot\a}M_{\dot\a \a}f^\a}
\end{equation}
where
\begin{equation}
\bar f^{\dot \a}M_{\dot\a\a}f^\a=\frac{1}{4}\bar f^{\dot
\a}(D_\a\bar D_{\dot\a}+\frac{3}{4}\bar D_{\dot\a}D_\a)f^\a
\end{equation}
and the gauge-fixing parameter is denoted by $\xi$. The gauge-fixing term becomes
\begin{equation}
S_{gf}=-\xi\int
{\mbox{d}}^8 z\hspace{1mm} (\bar D_{\dot\a}\bar\S)(\frac{3}{16}\bar
D^{\dot\a}D^{\a}+\frac 14 D^\a\bar D^{\dot\a})(D_\a\S) .
\end{equation}
One can easily show that the ghost fields are decoupled.

After the gauge-fixing, the part of the classical action 
quadratic in quantum superfields is given by
\begin{equation}
S^{(2)} = S^{(2)}_0 + S^{(2)}_{int}, \label{S2} 
\end{equation}
with
\begin{equation}
S^{(2)}_0=\frac{1}{2}\int {\mbox{d}}^8 z\hspace{1mm} 
\left(\begin{array}{cc}
\Sigma & \Sigma^{+} \end{array} \right){\mathcal
M} \left(\begin{array}{l}\Sigma\\
\Sigma^{+} \end{array} \right) \label{Snula}
\end{equation}
and
\begin{equation}
S^{(2)}_{int}=\frac{1}{2}\int {\mbox{d}}^8 z{\mbox{d}}^8 z'\hspace{1mm} 
\left(\begin{array}{cc}
\Sigma & \Sigma^{+} \end{array} \right)(z){\mathcal
V}(z,z') \left(\begin{array}{l}\Sigma\\
\Sigma^{+} \end{array} \right) (z'). \label{Sint}
\end{equation}
Kinetic and interaction terms are collected in the matrices $\cal{M}$ 
and $\cal{V}$ respectively. The matrix $\cal{M}$ is given by
\begin{equation}
\mathcal M= \left(\begin{array}{cc} -m\square^{1/2}P_{-} & \square
(P_2+\xi(P_1+P_T))\\ \square (P_1+\xi(P_2+P_T)) &
-m\square^{1/2}P_{+}
\end{array} \right) ,\label{M}
\end{equation}
with 
\begin{equation}
P_+ = \frac{D^2}{4\Box^{1/2}}, \quad  
P_- = \frac{{\bar D}^2}{4\Box^{1/2}}. \label{P+}
\end{equation}
The interaction matrix ${\cal V}$ is 
\begin{equation}
{\mathcal V}=\left(\begin{array}{cc} F & 0\\ 0 &
\bar{F}\end{array}\right) .\label{V}
\end{equation}
There are two types of elements in $\cal{V}$, local and nonlocal. We 
split them into $F_1$ and $F_2$
\begin{equation}
F(z,z')=F_{1}(z)\delta(z-z')+F_{2}(z,z'). \label{Fzz'}
\end{equation}
Elements of $F_1$ are given by
\begin{eqnarray}
F_{1} (z)&=& \sum_{i=0}^{10} F^{(i)}\nonumber\\
&=& -\frac{\lambda}{2} \Phi \bar{D}^2 -\frac{\lambda}{48}
K^{m}_{\ a}K^{*na} \overleftarrow{\bar{D}^{2} D^{\alpha}}
(\partial_{m}\Phi) \theta\theta\bar{\theta}\bar{\theta}
\partial_{n}D_{\alpha}\bar{D}^{2}\nonumber \\ 
&& - \frac{\lambda}{48} K^{m}_{\ a}K^{*na}
\overleftarrow{\bar{D}^{2} D^{\alpha}}
(\partial_{m}D_{\alpha}\Phi) \theta\theta\bar{\theta}\bar{\theta}
\partial_{n}\bar{D}^{2}\nonumber\\
&& - \frac{\lambda}{48} K^{m}_{\ a}K^{*na} \overleftarrow{\partial_{m}\bar{D}^{2}}
(D^{\alpha}\Phi) \theta\theta\bar{\theta}\bar{\theta}
\partial_{n}D_{\alpha}\bar{D}^{2} \nonumber\\
&&+\frac{\lambda}{24} K^{m}_{\ a}K^{*n}_{\ b}
\overleftarrow{\bar{D}^{2}D^{\alpha}}(\sigma^{ab})_{\alpha}^{\
\beta}(\partial_{m}\Phi)\theta\theta\bar{\theta}\bar{\theta}
\partial_{n} D_{\beta} \bar{D}^{2} \nonumber\\ & &+\frac{\lambda}{24} K^{m}_{\ a}K^{*n}_{\ b}
\overleftarrow{\partial_{m}\bar{D}^{2}}(D^{\alpha}\Phi)(\sigma^{ab})_{\alpha}^{\
\beta}\theta\theta\bar{\theta}\bar{\theta}
\partial_{n} D_{\beta} \bar{D}^{2} \nonumber\\ 
&&+\frac{\lambda}{24} K^{m}_{\ a}K^{*n}_{\ b}
\overleftarrow{\partial_{m}\bar{D}^{2}}(\partial_{n}D^{\alpha}\Phi)(\sigma^{ba})_{\alpha}^{\
\beta}\theta\theta\bar{\theta}\bar{\theta} D_{\beta} \bar{D}^{2}
\nonumber\\ 
&& -\frac{\lambda}{512} K^{mn}K_{mn}
\overleftarrow{\bar{D}^{2}D^{2}} \Phi \bar{\theta}\bar{\theta}
D^{2}\bar{D}^{2} \label{F1} \\ 
&& -\frac{\lambda}{96} K^{m}_{\
a}K^{*na} \overleftarrow{\partial_{m}\bar{D}^2} (\partial_{n}\Phi)
\theta\theta\bar{\theta}\bar{\theta} D^{2}\bar{D}^2 \nonumber\\ 
&& -\frac{\lambda}{192}K^{m}_{\ a}K^{*na}
\overleftarrow{\partial_{m}\bar{D}^{2}}
(D^{2}\Phi)\theta\theta\bar{\theta}\bar{\theta}
\partial_{n}\bar{D}^2\nonumber\\ 
&& +\frac{\lambda}{96} K^{m}_{\ a}K^{*na}
\overleftarrow{\square\bar{D}^{2}} \Big(\int {\mbox{d}}^8 z'\hspace{1mm}
(\partial_{m}D^{2}\Phi)(z') \frac{1}{\Box_{z'}}\delta(z'-z)\Big)
\theta\theta\bar{\theta}\bar{\theta}
\partial_{n}\bar{D}^2, \nonumber
\end{eqnarray}
while the elements of $F_2$ read
\begin{eqnarray}
F_{2}(z,z') &=& \sum_{i=11}^{12} F^{(i)}\nonumber\\
&=& \frac{\lambda}{96} K^{m}_{\ a}K^{*na}
\overleftarrow{\partial_{m}\bar{D}^{2} D^{2}} 
\frac{1}{\Box_{z'}}\delta(z'-z)\theta\theta\bar{\theta}\bar{\theta}
((\partial_{n}\Phi)\square\bar{D}^{2})(z')
\nonumber\\ 
&& + \frac{\lambda}{96} K^{m}_{\ a}K^{*na}
\overleftarrow{\partial_{m}\bar{D}^{2} D^{2}} 
\frac{1}{\Box_{z'}}\delta(z'-z)\theta\theta\bar{\theta}\bar{\theta}
(\Box\Phi \partial_{n}\bar{D}^{2})(z') .\label{F2'}
\end{eqnarray}

The one-loop effective action is then
\begin{equation}
\G=S_0+S_{int}+\frac{i}{2}\tr\log(1+{\mathcal M}^{-1}{\mathcal V}).
\label{ED-opsta}
\end{equation}
The last term in (\ref{ED-opsta}) is the one-loop correction to the
effective action and ${\cal M}^{-1}$ is the inverse of (\ref{M}) given by
\begin{equation}
{\mathcal M}^{-1}=\left(\begin{array}{cc} A & B\\ \bar{B} &
\bar{A} \end{array}\right)=\left(\begin{array}{cc}
\frac{mD^2}{4\square (\square-m^2)} &
\frac{D^2\bar{D}^2}{16\square (\square-m^2)}+\frac{\bar{D}^2 D^2-2\bar{D}D^2\bar{D}}{16\xi\square^2}\\
\frac{\bar{D}^2 D^2}{16\square
(\square-m^2)}+\frac{D^2\bar{D}^2-2D\bar{D}^2 D}{16\xi\square^2} &
\frac{m\bar{D}^2}{4\square (\square-m^2)}
\end{array}\right) .\label{Minv}
\end{equation}
Expansion of the logarithm in (\ref{ED-opsta}) leads to the one-loop 
corrections
\begin{equation}
\Gamma_1=\frac{i}{2}\tr\sum_{n=1}^{\infty}
\frac{(-1)^{n+1}}{n}({\mathcal
M}^{-1}\mathcal{V})^{n}=\sum^{\infty}_{n=1} \Gamma_{1}^{(n)}.
\label{1LoopExp}
\end{equation}

\section{One-point and two-point functions}

The first term in the expansion (\ref{1LoopExp}) gives the divergent part
of the one-point functions, the tadpole contribution. We obtain
\begin{equation}
\Gamma^{(1)}_{1}=\frac{i}{2} \tr ({\mathcal
M}^{-1}{\mathcal V})=\frac{i}{2} \tr
\big( AF+\bar{A}\bar{F}\big) = 0. \label{1point}
\end{equation}
Therefore just like in the commutative Wess-Zumino model
there is no tadpole contribution.

Next we calculate the divergent part of the two point functions. It is
given by
\begin{eqnarray}
\Gamma^{(2)}_{1} &=& -\frac{i}{4} \tr ({\mathcal
M}^{-1}{\mathcal V})^{2} \nonumber\\
&=& -\frac{i}{4} \tr \big( AFAF + 2
\bar{B}FB\bar{F} + \bar{A}\bar{F}\bar{A}\bar{F} \big) .\label{2point1}
\end{eqnarray}
First we calculate the $AFAF$ contributions. They are given by (remember
that $F^{(i)}$ is the i-th element of the expansions (\ref{F1}) and 
(\ref{F2'})) 
\begin{eqnarray}
{\rm Tr} (AF^{(0)}AF^{(0)})&=&0 ,\nonumber\\
{\rm Tr} (AF^{(1)}AF^{(0)})\bigg |_{d.p.}&=& -\frac{i m^{2}
\lambda^{2}K^{m}_{\ a}K^{*na}}{6\pi^{2}\varepsilon} \int 
{\mbox{d}}^4 x\hspace{1mm} 
\partial_{m}A\partial_{n}A ,\nonumber\\
{\rm Tr} (AF^{(2)}AF^{(0)})&=&0 ,\nonumber\\
{\rm Tr} (AF^{(3)}AF^{(0)})&=&0 ,\nonumber\\
{\rm Tr} (AF^{(4)}AF^{(0)})&=&0 ,\nonumber\\
{\rm Tr} (AF^{(5)}AF^{(0)})&=&0 ,\nonumber\\
{\rm Tr} (AF^{(6)}AF^{(0)})&=&0 ,\nonumber\\
{\rm Tr} (AF^{(7)}AF^{(0)})\bigg |_{d.p.}&=& - \frac{i
m^{2}\lambda^{2}
K^{mn}K_{mn}} {8\pi^{2}\varepsilon} \int {\mbox{d}}^4 x\hspace{1mm} 
F^{2}, \nonumber\\
{\rm Tr} (AF^{(8)}AF^{(0)})\bigg |_{d.p.}&=& \frac{i
m^{2}\lambda^{2}
K^{m}_{\ a}K^{*na}}{12\pi^{2}\varepsilon} \int {\mbox{d}}^4 x\hspace{1mm} 
\partial_{m}A\partial_{n}A ,\nonumber\\
{\rm Tr} (AF^{(9)}AF^{(0)}) &=&0 ,\nonumber\\
{\rm Tr} (AF^{(10)}AF^{(0)})&=&0 ,\nonumber\\
{\rm Tr} (AF^{(11)}AF^{(0)})\bigg |_{d.p.}&=&0 ,\nonumber\\
{\rm Tr} (AF^{(12)}AF^{(0)})\bigg |_{d.p.}&=&0.\nonumber
\end{eqnarray}
Adding these terms we obtain
\begin{eqnarray}
\tr (AFAF)\bigg |_{d.p.}&=& \tr (AF^{(0)}AF^{(0)})\bigg
|_{d.p.}+ 2\sum_{i=1}^{12}\tr (AF^{(i)}AF^{(0)})\bigg
|_{d.p.}\nonumber \\
&=& -\frac{i m^2 \lambda^2 K^{m}_{\ a}K^{*na}}{6\pi^2 \varepsilon} 
\int {\mbox{d}}^4 x\hspace{1mm} \partial_{m}A \partial_{n}A \nonumber \\
&& -\frac{i m^2 \lambda^2
K^{mn}K_{mn}}{4\pi^2\varepsilon}\int {\mbox{d}}^4 x\hspace{1mm} F^2.
\label{2pointA}
\end{eqnarray}

The $\bar{B}FB\bar{F}$ term is more difficult to calculate. Some of the
identities we use are given in the Appendix. We obtain the following 
contributions:
\begin{eqnarray}
\tr (\bar{B}F^{(0)}B\bar{F}^{(0)})\bigg |_{d.p.}
&=& \frac{i \lambda^{2}}{2\pi^{2}\varepsilon}\int 
{\mbox{d}}^8 z\hspace{1mm} \Phi^{\dagger}\Phi ,\nonumber\\
\tr (\bar{B}F^{(1)}B\bar{F}^{(0)})\bigg |_{d.p.}
&=& -\frac{i \lambda^{2}K^{m}_{\ a}K^{*na}}{12\pi^{2}\varepsilon}
\int {\mbox{d}}^4 x\hspace{1mm} A^*(\square - 4 m^2)\partial_{m}
\partial_{n} A ,\nonumber\\
\tr (\bar{B}F^{(2)}B\bar{F}^{(0)})\bigg |_{d.p.}&=& 
-\frac{\lambda^{2}K^{m}_{\ a}K^{*na}}{36\pi^{2}\varepsilon}
\int {\mbox{d}}^4 x\hspace{1mm} \bar{\psi} \bar{\sigma}^{l}\partial_{l}
\partial_{m}\partial_{n}\psi \nonumber \\
&& + \frac{\lambda^{2}K^{m}_{\ a}K^{*na}}{12\pi^{2}\varepsilon} 
\int {\mbox{d}}^4 x\hspace{1mm} \bar{\psi} \bar{\sigma}_{n} 
\bigg(m^2-\frac{\square}{6}\bigg)\partial_{m}\psi ,\nonumber\\
\tr (\bar{B}F^{(3)}B\bar{F}^{(0)})\bigg |_{d.p.}&=& 
\frac{\lambda^{2}K^{m}_{\ a}K^{*na}}{72\pi^{2}\varepsilon}
\int {\mbox{d}}^4 x\hspace{1mm} \bar{\psi} \bar{\sigma}^{l}\partial_{l}
\partial_{m}\partial_{n}\psi ,\nonumber\\
\tr (\bar{B}F^{(4)}B\bar{F}^{(0)})\bigg |_{d.p.}&=& 
-\frac{i \lambda^{2}K^{m}_{\ a}K^{*na}}{2\pi^{2}\varepsilon}
\int {\mbox{d}}^4 x\hspace{1mm} A^*\big(m^{2}-\frac{\square}{6}\big)
\partial_{m}\partial_{n} A ,\nonumber\\
\tr (\bar{B}F^{(5)}B\bar{F}^{(0)})\bigg |_{d.p.}&=&
-\frac{\lambda^{2}K^{m}_{\ a}K^{*n}_{\ b}}{72\pi^{2}\varepsilon}
\int {\mbox{d}}^4 x\hspace{1mm} \bar{\psi}(\bar{\sigma}^{b}\partial^{a}
- \bar{\sigma}^{a}\partial^{b} + i\varepsilon^{abcd}\bar{\sigma}_{d}
\partial_{c})\partial_{m}\partial_{n}\psi \nonumber\\
&&+ \frac{\lambda^2 K^{m}_{\ a}K^{*na}}{12\pi^{2}\varepsilon} 
\int {\mbox{d}}^4 x\hspace{1mm} \bar{\psi}\bar{\sigma}_{n} 
\bigg(m^2-\frac{\square}{6}\bigg)\partial_{m}\psi ,\nonumber\\
\tr (\bar{B}F^{(6)}B\bar{F}^{(0)})\bigg |_{d.p.}&=&
-\frac{\lambda^{2}K^{m}_{\ a}K^{*n}_{\ b}}{36\pi^{2}\varepsilon}
\int {\mbox{d}}^4 x\hspace{1mm} \bar{\psi}(\bar{\sigma}^{b}\partial^{a}
- \bar{\sigma}^{a}\partial^{b} + i\varepsilon^{abcd}\bar{\sigma}_{d}
\partial_{c})\partial_{m}\partial_{n}\psi \nonumber\\
&&- \frac{\lambda^2 K^{m}_{\ a}K^{*na}}{12\pi^{2}\varepsilon} 
\int {\mbox{d}}^4 x\hspace{1mm} \bar{\psi}\bar{\sigma}_{n} 
\bigg(m^2-\frac{\square}{6}\bigg)\partial_{m}\psi ,\nonumber\\
\tr (\bar{B}F^{(7)}B\bar{F}^{(0)})&=&0 ,\nonumber\\
\tr (\bar{B}F^{(8)}B\bar{F}^{(0)})\bigg |_{d.p.}&=& \frac{i m^{2}
\lambda^{2}K^{m}_{\ a}K^{*na}}{12\pi^{2}\varepsilon}
\int {\mbox{d}}^4 x\hspace{1mm} \partial_{m}A^*\partial_{n} A ,\nonumber\\
\tr (\bar{B}F^{(9)}B\bar{F}^{(0)})\bigg |_{d.p.}&=& 
\frac{i \lambda^{2}K^{m}_{\ a}K^{*na}}{72\pi^{2}\varepsilon}
\int {\mbox{d}}^4 x\hspace{1mm} F^*\partial_{m}\partial_{n}F ,\nonumber\\
\tr (\bar{B}F^{(10)}B\bar{F}^{(0)})\bigg |_{d.p.}&=& \frac{i m^{2}
\lambda^{2}K^{m}_{\ a}K^{*na}}{12\pi^{2}\varepsilon}
\int {\mbox{d}}^4 x{\mbox{d}}^4 y\hspace{1mm} \partial_{m}\partial_{n}
F(x)\square^{-1}_{x}\delta(x-y) F^*(y) ,\nonumber\\
\tr (\bar{B}F^{(11)}B\bar{F}^{(0)})\bigg |_{d.p.}&=& 
- \frac{i m^{2} \lambda^{2}K^{m}_{\ a}K^{*na}}{12\pi^{2}\varepsilon}
\int {\mbox{d}}^4 x\hspace{1mm} \partial_{m}A^*\partial_{n} A ,\nonumber\\
\tr (\bar{B}F^{(12)}B\bar{F}^{(0)})\bigg |_{d.p.}&=& 
-\frac{i \lambda^{2}K^{m}_{\ a}K^{*na}}{36\pi^{2}\varepsilon}
\int {\mbox{d}}^4 x\hspace{1mm} A^*\partial_{m}\partial_{n}\square A 
.\nonumber
\end{eqnarray}
Collecting all contributions we have
\begin{eqnarray}
\tr (\bar{B}FB\bar{F})\bigg |_{d.p.} &=& \tr
(\bar{B}F^{(0)}B\bar{F}^{(0)})\bigg |_{d.p.} 
+ 2 \sum_{i=1}^{12}\tr (\bar{B}F^{(i)}B\bar{F}^{(0)})\bigg |_{d.p.} 
\nonumber\\
&=&
\frac{i \lambda^{2}}{2\pi^{2}\varepsilon}\int {\mbox{d}}^8 z\hspace{1mm}
 \Phi^{+}\Phi \nonumber\\
&& - \frac{i \lambda^{2}K^{m}_{\ a}K^{*na}}{3\pi^{2}\varepsilon}
\int {\mbox{d}}^4 x\hspace{1mm} A^*\bigg(m^{2} 
+ \frac{\square}{6}\bigg)\partial_{m}\partial_{n} A \nonumber\\
&& - \frac{\lambda^2 K^{m}_{\ a}K^{*n}_{\ b}}{12\pi^{2}\varepsilon}
\int d^{4}x \bar{\psi}(\bar{\sigma}^{b}\partial^{a}-\bar{\sigma}^{a}
\partial^{b}+ i\varepsilon^{abcd}\bar{\sigma}_{d}\partial_{c})
\partial_{m}\partial_{n} \psi \nonumber \\
&& + \frac{\lambda^2 K^{m}_{\ a}K^{*na}}{6\pi^{2}\varepsilon}
\int {\mbox{d}}^4 x\hspace{1mm} \bar{\psi}\bar{\sigma}_{n}
\partial_{m}\bigg(m^{2}-\frac{\square}{6}\bigg)\psi \nonumber \\
&& - \frac{\lambda^2 K^{m}_{\ a}K^{*na}}{36\pi^{2}\varepsilon}
\int {\mbox{d}}^4 x\hspace{1mm}
\bar{\psi}\bar{\sigma}^{l}\partial_{l}\partial_{m}\partial_{n} \psi \nonumber \\
&& + \frac{i \lambda^{2}K^{m}_{\
a}K^{*na}}{36\pi^{2}\varepsilon} \int {\mbox{d}}^4 x\hspace{1mm}
F^*\partial_{m}\partial_{n} F \label{2pointB}\\ 
&& +\frac{i m^{2}\lambda^{2}K^{m}_{\
a}K^{*na}}{6\pi^{2}\varepsilon} 
\int {\mbox{d}}^4 x {\mbox{d}}^4 y\hspace{1mm}
\partial_{m}\partial_{n}F(x) \square^{-1}_{x}\delta(x-y)
F^*(y). \nonumber
\end{eqnarray}
Finally, adding (\ref{2pointA}) and (\ref{2pointB}) we obtain the 
divergent part of the two-point function
\begin{eqnarray}
\Gamma^{(2)}_{1}\bigg|_{d.p.}&=& 
 -\frac{m^2 \lambda^2 K^{m}_{\ a}K^{*na}}{24\pi^2 \varepsilon}
\int {\mbox{d}}^4 x\hspace{1mm} (\partial_{m}A \partial_{n}A 
+ \partial_{m}A^* \partial_{n}A^*) \nonumber \\
&& -\frac{m^2 \lambda^2
K^{mn}K_{mn}}{16\pi^2\varepsilon}\int {\mbox{d}}^4 x\hspace{1mm}
 F^2 \nonumber\\ 
&& -\frac{m^2 \lambda^2
K^{*mn}K^{*}_{mn}}{16\pi^2\varepsilon}
\int {\mbox{d}}^4 x\hspace{1mm} F^{*2} \nonumber\\
&& +\frac{\lambda^{2}}{4\pi^{2}\varepsilon}
\int {\mbox{d}}^8 z\hspace{1mm} \Phi^{+}\Phi \nonumber\\
&& - \frac{\lambda^{2}K^{m}_{\ a}K^{*na}}{6\pi^{2}\varepsilon}
\int {\mbox{d}}^4 x\hspace{1mm} A^*\bigg(m^{2}
+ \frac{\square}{6}\bigg)\partial_{m}\partial_{n} A \nonumber\\
&& + \frac{i\lambda^2 K^{m}_{\ a}K^{*n}_{\ b}}{24\pi^{2}\varepsilon}
\int {\mbox{d}}^4 x\hspace{1mm} \bar{\psi}(\bar{\sigma}^{b}\partial^{a}
-\bar{\sigma}^{a}\partial^{b}+ i\varepsilon^{abcd}\bar{\sigma}_{d}
\partial_{c})\partial_{m}\partial_{n} \psi \nonumber \\
&& - \frac{i\lambda^2 K^{m}_{\ a}K^{*na}}{12\pi^{2}\varepsilon}
\int {\mbox{d}}^4 x\hspace{1mm}
\bar{\psi}\bar{\sigma}_{n}
\partial_{m}\bigg(m^{2}-\frac{\square}{6}\bigg)\psi \nonumber \\
&& + \frac{i\lambda^2 K^{m}_{\ a}K^{*na}}{72\pi^{2}\varepsilon} 
\int {\mbox{d}}^4 x\hspace{1mm}
\bar{\psi}\bar{\sigma}^{l}\partial_{l}\partial_{m}\partial_{n} \psi 
\nonumber \\
&& + \frac{\lambda^{2}K^{m}_{\ a}K^{*na}}{72\pi^{2}\varepsilon}
\int {\mbox{d}}^4 x\hspace{1mm}
F^*\partial_{m}\partial_{n} F \nonumber\\ 
&& + \frac{m^{2}\lambda^{2}K^{m}_{\ a}K^{*na}}{12\pi^{2}\varepsilon}
\int {\mbox{d}}^4 x {\mbox{d}}^4 y\hspace{1mm}
\partial_{m}\partial_{n}F(x) \square^{-1}_{x}\delta(x-y)
F^*(y). \label{2pointAll}
\end{eqnarray}
We immediately see that the divergences appearing in (\ref{2pointAll}) cannot
be absorbed by counterterms since the terms appearing in (\ref{2pointAll})
do not exist in the classical action. All terms in (\ref{2pointAll})
quadratic in the deformation parameter are also quadratic in fields. 
However, the deformation of the classical action (\ref{SGaction}) is only present in the interaction term, and terms in the action quadratic in the deformation
parameter will always be of the third order in fields. We have to conclude that
our model, as it stands, is not renormalizable. 

\section{Discussion and conclusions}

Let us now summarize what we have done so far and discuss the obtained 
results in more detail.

In order to see how different deformations (different twists) affect 
renormalizability of 
the Wess-Zumino model, we considered one special
example of twist, (\ref{twist}). The main adventage of this twist
is that it is hermitian and therefore implies the hermitian $\star$-product. 
Compared with the undeformed
SUSY Hopf algebra, the twisted SUSY Hopf algebra changes. In
particular, the Leibniz rule (\ref{deftrlaw}) becomes deformed.
The notion of chirality is lost and we had to apply the
method of projectors introduced in \cite{miSUSY} to obtain the
action. A nonlocal deformation of the commutative Wess-Zumino action 
invariant under the deformed SUSY transformations (\ref{defsusytr}) and with a good commutative limit was introduced and its 
renormalizability properties were investigated. Notice that the nonlocality
comes from the application of the chiral projector $P_2$\footnote{Unlike
the Moyal-Weyl $\star$-product, the $\star$-product (\ref{star}) is finite and it does not 
introduce non-locality.}.

To calculate the divergent part of the effective action we used the
background field method and the supergraph technique. Like in the 
commutative Wess-Zumino model, there is no tadpole contribution. There is no mass counterterm which is again the same as in the undeformed Wess-Zumino model.
However, the divergent part of the two-point function cannot be canceled
and we have to conclude that our model is not renormalizable. Calculating
divergent parts of the three-point and higher functions does not make sense
and it is technically very demanding.

Having in mind results of \cite{on-shell}, we also investigated on-shell  
renormalizability of our model. In general, on-shell renormalizability 
leads to a one-loop renormalizable $S$-matrix. On the other hand, 
one-loop on-shell renormalizable Green functions may spoil 
renormalizability at higher loops. After using the equations of motion which follow form the action (\ref{SGaction}) to obtain the on-shell 
divergent terms we see that the divergences in 
the two-point function remain and therefore the model is also not on-shell renormalizable.

In our previous work \cite{D-def-us} we had a similar problem, a deformed model which was not renormalizable. To obtain a renormalizable model we had to relax the condition of minimality of deformation and to include non-minimal terms. Also, in \cite{Penati} new terms of the form 
$\int {\mbox{d}}^8 z \theta\theta\bar{\theta}\bar{\theta} D^2 \Phi$ and
$\int {\mbox{d}}^8 z \theta\theta\bar{\theta}\bar{\theta} (D^2 \Phi)^2$ were
added in order to absorb divergences produced by 
$\int {\mbox{d}}^4 x F^3 
= \int {\mbox{d}}^8 z \theta\theta\bar{\theta}\bar{\theta} (D^2 \Phi)^3$ 
term. Since the model we work with is more complicated than the models of
\cite{D-def-us} and \cite{Penati}, it is not obvious which terms should
be added. Let us list possible terms. Note that the new terms 
have to be invariant under the deformed SUSY transformations 
(\ref{defsusytr}). This requirement gives three possibilities:
\begin{eqnarray}
T_1 &=& \int {\mbox{d}}^4 x\hspace{1mm} 
P_1(\Phi\star\Phi)\Big|_{\bar{\theta}\bar{\theta}} 
\label{T1}\\
&=& \frac{1}{2} K^{ab}K_{ab}\int {\mbox{d}}^4 x\hspace{1mm} \big( 
\frac{1}{2}(\p_m\psi)\sigma^m\bar{\sigma}^n(\p_n\psi)-F\Box A
\big). \nonumber\\
T_2 &=& \int {\mbox{d}}^4 x\hspace{1mm} P_1(\Phi\star P_2(\Phi\star\Phi))\Big|_{\bar{\theta}\bar{\theta}} 
\label{T2}\\
&=& \frac{1}{4}K^{ab}K_{ab}\int {\mbox{d}}^4 x\hspace{1mm} \big(
-AF \Box A - \frac{1}{2}F\Box A^2 \nonumber\\
&& + \frac{1}{2}\psi\psi\Box A +
\p_m(A\psi)\sigma^m\bar{\sigma}^n(\p_n\psi)  \big).
\nonumber\\
T_3 &=& \int {\mbox{d}}^4 x\hspace{1mm} \Phi\star P_1(\Phi\star\Phi))
\Big|_{\theta\theta\bar{\theta}\bar{\theta}} 
\label{T3}\\
&=& \frac{3}{4}K^{ab}K_{ab} \int {\mbox{d}}^4 x\hspace{1mm} \Big(
F(\p_m\psi)\sigma^m\bar{\sigma}^l(\p_l\psi) - F^2\Box A \Big)\nonumber\\
&& + K^{m}_{\ a}K^{*na} \int {\mbox{d}}^4 x\hspace{1mm} \Big(
A(\Box A)(\p_m\p_n A) + A(\p_m\p_l A)(\p_n\p^l A)
\Big). \nonumber
\end{eqnarray}
The term $T_1$ produces divergences of the type 
$\int {\mbox{d}}^4 x\hspace{1mm} \Phi^+ \Phi$ so it would not
spoil the renormalizability of the model. However, it cannot improve renormalizability
since divergences appearing in (\ref{2pointAll}) are not of the type
$T_1$. The term $T_2$ produces additional divergences that cannot be 
absorbed, so we have to ignore it. The $T_3$ term does not cancel
any of the terms present in the action (\ref{Sincompfields}). 
Additionally, it produces new divergent terms. However these terms might
look like, they can never cancel all the divergences in (\ref{2pointAll}) as divergences proportional to $K^{m}_{\ a}K^{*n}_{\ b}$ will
remain. This analysis forces us to conclude that even with a non-minimal
deformation our model remains nonrenormalizable.

Let us make a remark about the nonrenormalization theorem and its modifications in the case of deformed SUSY. One easily sees that the divergent terms of the effective action (\ref{2pointAll}) can be rewritten as
\begin{equation}
\Gamma^{(2)}_1\bigg|_{d.p.} = \int {\mbox{d}}^4 x_1 \hspace{1mm} {\mbox{d}}^4 x_2\hspace{1mm} {\mbox{d}}^2 \theta {\mbox{d}}^2 \bar{\theta}\hspace{1mm}
G_2(x_1, x_2, U_{(1)}, U_{(2)}) f_1(x_1, \theta, \bar{\theta}) f_2(x_2, \theta, \bar{\theta}),
\label{NRTh}
\end{equation}
with $f_i = f_i (\Phi, \Phi^+, D\Phi, \bar{D}\Phi, D\Phi^+, \bar{D}\Phi^+,\dots)$. The nonlocal term in (\ref{NRTh}) appears as a consequence of nonlocality in the classical action
(\ref{SGaction}). The result (\ref{NRTh}) confirms the modified nonrenormalization theorem \cite{Penati}.
The appearance of the spurion fields in (\ref{NRTh}) signals breaking of the undeformed SUSY. In our case, symmetry which remains after the breaking is the twisted SUSY (\ref{defsusytr}). However, it seems that the twisted SUSY is not enough to guarantee renormalizability.

It is obvious that different deformations obtained from 
different twists lead to models
with different quantum properties. In our previous work 
\cite{D-def-us} we studied a 
deformation which preserves the full undeformed SUSY. There, after 
relaxing the condition of minimality of deformation, we obtained a 
renormalizable Wess-Zumino model. In this paper we work with a
deformation given in terms of the non-SUSY-covariant derivatives. The 
Leibniz rule for the SUSY transformation (\ref{defsusytr}) changes and the 
deformed action (\ref{Sincompfields}) though invariant under twisted SUSY transformations,
is not invariant under the undeformed SUSY transformations. For example, the term $K^{mn}K_{mn}F^3$ breaks the undeformed SUSY. On the other hand, the twisted SUSY allows this term as a part of the invariant term $P_2(\Phi\star P_2(\Phi\star \Phi))\Big|_{\theta\theta}$, see Equations (5.13) and (5.14) in \cite{miSUSY}. 

The classical properties of theories with twisted symmetries are not fully understood \cite{NCbookMi}, \cite{TwistedNoeth}. For example, one cannot apply standard methods to find conserved charges and the modification of Noether theorem in the case of twisted symmetry has not been formulated yet. In this paper we analyze quantum properties of the theory with the twisted SUSY. This is the first time that renormalizability of a theory with a twisted symmetry is analyzed. Even after relaxing
the condition of minimality of deformation our model remains nonrenormalizable. This indicates that theories with twisted symmetries do not have the same quantum properties as theories with undeformed symmetries. In our example, we see that the twisted SUSY is not enough to guarantee renormalizability of the Wess-Zumino model. It is obvious that a better understanding of the twisted symmetry and its consequences, both classical and quantum is needed.

\appendix

\section{Calculation}

In this appendix we collect details of some calculations and some important
side results.

\begin{itemize}

\item Transformation laws of the component fields of the superfield 
${\rm F}$
(\ref{F}): 
\begin{eqnarray}
\delta_\xi f &=& \xi^\alpha \phi_\alpha
+ \bar{\xi}_{\dot\alpha}\bar{\chi}^{\dot\alpha}, \label{susytrf} \\
\delta_\xi \phi_\alpha &=& 2\xi_\alpha m
+ \sigma^m_{\ \alpha\dot{\alpha}}\bar{\xi}^{\dot\alpha}\big( v_m + i (\p_m f) \big), \label{susytrphi} \\
\delta_\xi \bar{\chi}^{\dot\alpha} &=& 2\bar{\xi}^{\dot\alpha}n
+ \bar{\sigma}^{m\dot{\alpha}\alpha}\xi_\alpha\big( -v_m + i (\p_m f) \big), \label{susytrchi} \\
\delta_\xi m &=&\bar{\xi}_{\dot\alpha}\bar{\lambda}^{\dot\alpha}
+ \frac{i}{2}\bar{\xi}_{\dot\alpha}\bar{\sigma}^{m\dot{\alpha}\alpha}(\p_m \phi_\alpha) , \label{susytrm} \\
\delta_\xi n &=& \xi^\alpha \varphi_\alpha
+ \frac{i}{2}\xi^\alpha\sigma^m_{\ \alpha\dot{\alpha}}(\p_m \bar{\chi}^{\dot\alpha}), \label{susytrn} \\
\sigma^m_{\ \alpha\dot{\alpha}}\delta_\xi v_m &=&
-i(\p_m\phi_\alpha)\xi^\beta\sigma^m_{\ \beta\dot{\alpha}} +
2\xi_\alpha\bar{\lambda}_{\dot\alpha} \nonumber\\
&&+ i\sigma^m_{\
\alpha\dot{\beta}}\bar{\xi}^{\dot\beta}(\p_m
\bar{\chi}_{\dot\alpha})
+ 2\varphi_\alpha\bar{\xi}_{\dot\alpha} , \label{susytrvm} \\
\delta_\xi \bar{\lambda}^{\dot\alpha} &=& 2\bar{\xi}^{\dot\alpha}d
+ i\bar{\sigma}^{l\dot{\alpha}\alpha}\xi_\alpha(\p_l m)
+ \frac{i}{2}\bar{\sigma}^{l\dot{\alpha}\alpha}\sigma^m_{\ \alpha\dot{\beta}}\bar{\xi}^{\dot\beta}(\p_m v_l) , \label{susytrlambda} \\
\delta_\xi \varphi_\alpha &=& 2\xi_\alpha d + i\sigma^l_{\
\alpha\dot{\alpha}}\bar{\xi}^{\dot\alpha}(\p_l n)
-\frac{i}{2}\sigma^l_{\ \alpha\dot{\alpha}}\bar{\sigma}^{m\dot{\alpha}\beta}\xi_\beta(\p_m v_l) , \label{susytrvarphi} \\
\delta_\xi d &=& \frac{i}{2}\xi^\alpha \sigma^m_{\
\alpha\dot{\alpha}}(\p_m \bar{\lambda}^{\dot\alpha})
-\frac{i}{2}(\p_m \varphi^\alpha)\sigma^m_{\
\alpha\dot{\alpha}}\bar{\xi}^{\dot\alpha} . \label{susytrd}
\end{eqnarray}

\item Irreducible components the superfield ${\rm F}$:
\begin{eqnarray}
P_2 {\rm F} &=& \frac{1}{16} \frac{\bar{D}^2 D^2 }{\Box} {\rm F} \nonumber\\
&=& \frac{1}{\Box}\Big( d - \frac{i}{2}(\p_m v^m) +
\frac{1}{4}\Box f\Big) + \sqrt{2}\theta^\alpha\Big(
\frac{i}{\sqrt{2}\Box}\sigma^m_{\ \alpha\dot{\alpha}}
(\p_m \bar{\lambda}^{\dot{\alpha}}) + \frac{1}{2\sqrt{2}}\phi_\alpha \Big) \nonumber\\
&& + \theta\theta m
+ i\theta\sigma^l \bar{\theta}\p_l\Big( \frac{d}{\Box} - \frac{i}{2\Box}(\p_m v^m)
+ \frac{1}{4}f \Big) \label{P2F'}\\
&& + \frac{1}{\sqrt{2}}\theta\theta\bar{\theta}_{\dot\alpha}\Big( \frac{1}{\sqrt{2}}\bar{\lambda}^{\dot{\alpha}}
+ \frac{i}{2\sqrt{2}}\bar{\sigma}^{m\dot{\alpha}\alpha}(\p_m\phi_\alpha) 
\Big) \nonumber\\
&& +\frac{1}{4}\theta\theta\bar{\theta}\bar{\theta}\Big(
d - \frac{i}{2}(\p_m v^m) + \frac{1}{4}\Box f \Big) .\nonumber
\end{eqnarray}
\begin{eqnarray}
P_1 {\rm F}&=& \frac{1}{16} \frac{D^2\bar{D}^2}{\Box} {\rm F} \nonumber\\
&=& \frac{1}{\Box}\Big( d + \frac{i}{2}(\p_m v^m) +
\frac{1}{4}\Box f\Big) + \sqrt{2}\bar{\theta}_{\dot\alpha}\Big(
\frac{i}{\sqrt{2}\Box}\bar{\sigma}^{m\dot{\alpha}\alpha}
(\p_m \varphi_\alpha) + \frac{1}{2\sqrt{2}}\bar{\chi}^{\dot\alpha} \Big) \nonumber\\
&& + \bar{\theta}\bar{\theta} n
- i\theta\sigma^l \bar{\theta}\p_l\Big( \frac{d}{\Box} + \frac{i}{2\Box}(\p_m v^m)
+ \frac{1}{4}f \Big) \label{P1F'}\\
&& + \frac{1}{\sqrt{2}}\bar{\theta}\bar{\theta}\theta^\alpha\Big( 
\frac{1}{\sqrt{2}}\varphi_\alpha
+ \frac{i}{2\sqrt{2}}\sigma^m_{\ \alpha\dot{\alpha}}(\p_m\bar{\chi}^{\dot\alpha}) 
\Big) \nonumber\\
&&+\frac{1}{4}\theta\theta\bar{\theta}\bar{\theta}\Big(
d + \frac{i}{2}(\p_m v^m) + \frac{1}{4}\Box f \Big) ,\nonumber
\end{eqnarray}
\begin{eqnarray}
P_T {\rm F} &=& \frac{1}{2}f - \frac{2}{\Box}d +\theta^\alpha\Big(
\frac{1}{2}\phi_\alpha-
i\frac{1}{\Box}\sigma^m_{\ \alpha\dot{\alpha}}\p_m\bar{\lambda}^{\dot\alpha}\Big) \nonumber\\
&& +\bar\theta_{\dot{\alpha}}\Big( \frac{1}{2}\bar\chi^{\dot{\alpha}}-i\frac{1}{\Box}\bar{\sigma}^{m\dot{\alpha}\alpha}
\p_m\varphi_{\alpha} \Big)
+\theta\sigma^m\bar{\theta} \Big( v_m-\frac{1}{\Box}\p_m\p_l v^l \Big)\nonumber\\
&& +\theta\theta\bar{\theta}_{\dot{\alpha}}\Big( \frac{1}{2}
\bar{\lambda}^{\dot\alpha} - \frac{i}{4}\bar{\sigma}^{m\dot{\alpha}\alpha}(\p_m\phi_\alpha)\Big)
+ \bar{\theta}\bar{\theta}\theta^\alpha\Big( \frac{1}{2}
\varphi_\alpha - \frac{i}{4}\sigma^m_{\ \alpha\dot{\alpha}}(\p_m\bar{\chi}^{\dot\alpha})\Big)\nonumber\\
&& +\frac{1}{4}\theta\theta\bar{\theta}\bar{\theta} \Big( 2d - \frac{1}{2}\Box f\Big) .\label{PTF}
\end{eqnarray}
The following identity holds
\begin{equation}
P_T = I-P_1-P_2 .\label{PT'}
\end{equation}

\item Some general formulas for the divergent parts of traces, where $K=\Box-m^2$
\begin{eqnarray}
\tr(K^{-1}f) &=& \frac{i}{8\pi^2\epsilon}m^2\int{\mbox{d}}^4
x\hspace{1mm} f ,\label{f1} \\
\tr(\partial_aK^{-1}f) &=&0 , \label{f2}\\
\tr(\Box K^{-1}f) &=& \frac{im^4}{8\pi^2\epsilon}\int{\mbox{d}}^4 x
\hspace{1mm} f ,\label{f3}\\
\tr(\Box^2K^{-1}f) &=& \frac{im^6}{16\pi^2\epsilon}\int{\mbox{d}}^4
x\hspace{1mm} f  , \label{f4}\\
\tr(K^{-1}fK^{-1}g)&=&\frac{i}{8\pi^2\epsilon}\int{\mbox{d}}^4
x\hspace{1mm} fg ,\label{f5}\\
\tr(\partial_nK^{-1}fK^{-1}g) &=&\frac{i}{16\pi^2\epsilon}
\int{\mbox{d}}^4
x\hspace{1mm} \partial_nfg ,\label{f6}\\
\tr(\partial_nK^{-1}f\partial_mK^{-1}g) &=& -\frac{i}{16\pi^2\epsilon}
\int{\mbox{d}}^4 x\hspace{1mm} \label{f7}\\
&&\hspace{-4cm}f\Big( \frac{1}{3} \partial_n\partial_m+\frac{1}{6}\eta_{mn}\Box
-\eta_{mn}m^2\Big)g , \nonumber\\
\tr(\partial_nK^{-1}f\partial_m\partial_pK^{-1}g) &=& -\frac{i}{32\pi^2\epsilon}
\int{\mbox{d}}^4 x\hspace{1mm} \label{f8}\\
&&\hspace{-4cm}f\Big( \frac{1}{3} \partial_n\partial_m\partial_p+(\eta_{mp}\partial_{n}-\eta_{np}\partial_{m}- \eta_{nm}\partial_{p})(m^{2}-\frac{1}{6}\Box)\Big)g . \nonumber
\end{eqnarray}
\end{itemize}

\vspace*{0.5cm}
\begin{flushleft}
{\Large {\bf Acknowledgments}}
\end{flushleft}

The work of the authors is supported by the
project $141036$ of the Serbian Ministry of Science.

\end{document}